\documentclass[11pt]{article}
\usepackage{fullpage}
\usepackage{cite}
\usepackage{graphicx}
\usepackage{amsmath}
\usepackage{amssymb}
\usepackage{subcaption}
\usepackage{mathtools}
\usepackage{xcolor}
\usepackage[most]{tcolorbox}
\usepackage{blindtext}
\usepackage{tikz-cd}
\usepackage{subfiles}
\usepackage{titlesec}
\usepackage{parskip}
\title{}
\date{}

\def\beq{\begin{equation}}
\def\eeq{\end{equation}}
\allowdisplaybreaks
\begin{document}
\bibliographystyle{utphys}
\newcommand{\sech}{sech}
\newcommand{\csch}{csch}

\newcommand{\Ft}{\tilde{F}}
\newcommand{\msbar}{\ensuremath{\overline{\text{MS}}}}
\newcommand{\DIS}{\ensuremath{\text{DIS}}}
\newcommand{\abar}{\ensuremath{\bar{\alpha}_S}}
\newcommand{\bb}{\ensuremath{\bar{\beta}_0}}
\newcommand{\rc}{\ensuremath{r_{\text{cut}}}}
\newcommand{\Nd}{\ensuremath{N_{\text{d.o.f.}}}}
\setlength{\parindent}{0pt}

\titlepage
\begin{flushright}
QMUL-PH-20-12
\end{flushright}

\vspace*{0.5cm}

\begin{center}
{\bf \Large Weyl doubling}

\vspace*{1cm} \textsc{Rashid Alawadhi\footnote{r.alawadhi@qmul.ac.uk},
David S. Berman\footnote{d.s.berman@qmul.ac.uk},
Bill Spence\footnote{w.j.spence@qmul.ac.uk}  } \\

\vspace*{0.5cm} Centre for Research in String Theory\\ School of
Physics and Astronomy \\
Queen Mary University of London \\
London E1 4NS UK\\

\end{center}

\vspace*{0.5cm}

\begin{abstract}

We study a host of spacetimes where the Weyl curvature may be expressed algebraically in terms of an Abelian field strength. These include Type D spacetimes in four and higher dimensions which obey a simple quadratic relation between the field strength and the Weyl tensor, following the Weyl spinor double copy relation. However, we  diverge from the usual double copy paradigm by taking the gauge fields to be in the curved spacetime as opposed to an auxiliary flat space.

We show how for Gibbons-Hawking spacetimes with more than two centres a generalisation of the Weyl doubling formula is needed by including a  derivative-dependent expression which is linear in the Abelian field strength. We also find a type of twisted doubling formula in a case of a manifold with Spin(7) holonomy in eight dimensions. 

For Einstein Maxwell theories where there is an independent gauge field defined on spacetime, we investigate how the gauge fields determine the Weyl spacetime curvature via a doubling formula. We first  show that this occurs for the Reissner-Nordstr\"om metric in any dimension, and that this generalises to the electrically-charged Born-Infeld solutions. Finally, we consider brane systems in supergravity, showing  that a similar doubling formula applies. This Weyl formula is based on the field strength of the p-form potential that minimally couples to the brane and the brane world volume Killing vectors.

\end{abstract} 

\vspace*{0.5cm}

\section{Introduction}
\label{sec:intro}

The ``double copy" and its inverse the ``single copy" have received considerable attention in the past few years, as they provide an intriguing link between gauge theories and gravity. The double copy refers to moving from gauge theory to gravity while the single copy is the inverse map (there is also a ``zeroth copy" to a scalar theory).
This relationship, as a map between perturbative scattering amplitudes in gauge theory and gravity, was first studied in\cite{Bern:2008qj,Bern:2010ue,Bern:2010yg}. A tree-level proof has been given~\cite{BjerrumBohr:2009rd,Stieberger:2009hq,Bern:2010yg,BjerrumBohr:2010zs,Feng:2010my,Tye:2010dd,Mafra:2011kj,Monteiro:2011pc,BjerrumBohr:2012mg},
where it has a stringy origin~\cite{Kawai:1985xq}, and support for its existence at loop level found in\cite{Bern:2010ue,Bern:1998ug,Green:1982sw,Bern:1997nh,Carrasco:2011mn,Carrasco:2012ca,Mafra:2012kh,Boels:2013bi,Bjerrum-Bohr:2013iza,Bern:2013yya,Bern:2013qca,Nohle:2013bfa,Bern:2013uka,Naculich:2013xa,Du:2014uua,Mafra:2014gja,Bern:2014sna,Mafra:2015mja,He:2015wgf,Bern:2015ooa, Mogull:2015adi,Chiodaroli:2015rdg,Bern:2017ucb,Johansson:2015oia,Oxburgh:2012zr,White:2011yy,Melville:2013qca,Luna:2016idw,Saotome:2012vy,Vera:2012ds,Johansson:2013nsa,Johansson:2013aca}.

Subsequently the double/single copy was applied to some exact classical solutions. The Schwarzschild solution was shown to single copy to an electric charge \cite{Monteiro:2014cda}, the Taub-NUT solution to a magnetic monopole \cite{Luna:2015paa} and the  Eguchi-Hanson solution  mapped to a self-dual gauge field \cite{Berman:2018hwd}, for example. More general topologically non-trivial solutions have been double copied in the work of \cite{Sabharwal:2019ngs}. Other work examining symmetries of the linearised double copy is \cite{Anastasiou:2014qba,Borsten:2015pla,Anastasiou:2016csv,Anastasiou:2017nsz,Cardoso:2016ngt,Borsten:2017jpt,Anastasiou:2017taf,Anastasiou:2018rdx,LopesCardoso:2018xes}. More recent work has developed a wide variety of analysis applied to the double copy \cite{Carrillo-Gonzalez:2018pjk} -\cite{Adamo:2020qru}. 

In \cite{Luna:2018dpt}  four-dimensional type D spacetimes were investigated, using a double copy formula for the Weyl curvature spinor in terms of a Maxwell spinor. 
Duality symmetries of gauge theories and their relationship to solution-generating maps in gravity have also been studied recently from the point of view of the double copy \cite{Alawadhi:2019urr,Huang:2019cja}. The earlier work of \cite{Mars:2001gd} had used a self-dual Maxwell field, defined in terms of a Killing vector on the spacetime, in order to study how the Weyl tensor transformed under $sl(2,R)$, and noted in particular that if the Weyl tensor was given by a suitable function quadratic in the Maxwell field, then the $sl(2,R)$-transformed metric also had a Weyl tensor satisfying this property. In \cite{Alawadhi:2019urr} we, with Peinador Veiga,  studied various metrics for which this is the case, showing how they transform under duality. This work, and that of  \cite{Luna:2018dpt}, suggested that it would be interest to study cases where the Weyl tensor is given in terms of an Abelian gauge field, by what we will call ``Weyl doubling". 

In this paper we would thus like to explore classes of gravity and gravity-gauge field systems where the Weyl curvature of a spacetime is given as a quadratic function of an Abelian field strength which is also defined on the spacetime. Crucially we differ from the normal double copy scheme as our gauge field will be taken to be in the curved space time rather than some auxiliary flat background.
We will investigate two such classes - the first where the gauge  field is defined using intrinsic geometric properties of the spacetime, and the second where it is an additional field in the theory.  

This paper is organised as follows.
In Section 2 we introduce the definition of purely algebraic Weyl doubling. We then show that the spacetimes obeying Weyl doubling satisfy the conditions for type D in four and higher dimensions. The independent scalar curvature invariants are then functions of the invariants constructed from the field strength, and  the Weyl-NP scalars are in turn functions of the Maxwell-NP scalars. We give a number of examples.

In Section 3 we then study gravitational instantons where the curvature satisfies a self-duality condition. For the Gibbons-Hawking metrics, the anti-self-dual two form field strength derived from the Killing vector is expanded in the triplet of anti-self-dual two forms and we show that these coefficients determine the Weyl tensor of the metric via two terms - the first being a direct tensor doubling formula as in Section 2, and the second involving derivatives of the Abelian field strength. Based on this we give a derivative correction to the Weyl doubling formula. 
In higher dimensions, we find a  manifold with Spin(7) holonomy that obeys a doubling construction using a type of twisted Maxwell field.  

We next consider  examples where there is an independent gauge field defined on the spacetime and investigate if the gauge fields in these gravity-gauge systems obey a doubling formula for the Weyl tensor. In Section 4 we first  give an example for the case of the Reissner-Nordstr\"om metric in any dimension, showing that the  Weyl tensor is given by a doubling formula based on the external electromagnetic field strength. This formula  applies to both the extremal and the non-extremal Reissner-Nordstr\"om solutions. We show that this generalises to the charged Born-Infeld solution in any dimension. In Section 5 we then consider brane systems in supergravity. Here we show that the components of the Weyl tensor are given by a simple doubling  formula based on the field strength of the p-form that minimally couples to the brane world volume and the brane Killing vectors. We finish with a discussion of the results and future work in Section 6.


\section{Weyl doubling}
\label{sec:Weyld}

\subsection{Spacetime classification}
The central object of study will be the ``doubling" formula, where a tensor, $C[F]$ with the algebraic symmetries of the Weyl tensor in $D$ dimensions can be constructed from an $n$-form $F=dA$ (with the gauge field $A$ an $n-1$ form) as follows:
\begin{equation}
\label{Dhigher}
C_{\mu\nu\rho\sigma} [F] = F_{\mu\nu}\cdot F_{\rho\sigma} - F_{\rho\mu}\cdot F_{\nu\sigma}-\frac{6}{D-2}g_{\mu\rho}{F_\nu} \cdot F_{\sigma} + \frac{3}{(D-1)(D-2)}g_{\mu\rho}g_{\nu\sigma}F\cdot F   \; \Big\vert_s  \, ,
\end{equation}
where a dot product  means to contract all non-visible indices between the two terms, for example for an $n$-form,
$F_{\mu\nu}\cdot F_{\rho\sigma} = F_{\mu\nu{\lambda_3}\dots{\lambda_n}} {F_{\mu\nu}}^{{\lambda_3}\dots{\lambda_n}}$.
The symbol ``$\vert_s$" above applies to the expression on the right-hand side of the equation, and it means to anti-symmetrise in the indices $\mu,\nu$ and in $\rho, \sigma$, with unit weight. 

We wish to study cases where the spacetime Weyl tensor $C_{\mu\nu\rho\pi}[g]$  is proportional to this expression - 
\begin{equation}
\label{CeqCF}
C_{\mu\nu\rho\pi}[g] = \frac{1}{\sigma}\, C_{\mu\nu\rho\pi} [F]  \, 
\end{equation}
for some function $\sigma$.  In a later example we will exhibit a derivative corrected expression where the right-hand side of \eqref{CeqCF} contains further terms determined by derivatives of $F$.
The field strength $F$ is closed and required to be divergence-free. (Note that the field is defined on the spacetime with metric $g_{\mu \nu}$ thus the divergence equation is non-trivial.)
\par

Let us first consider the case $n=2$ where we have a normal Abelian two form field strength. (The cases with $n>2$ will be relevant when we discuss supergravity brane solutions.) One can identify two cases where the divergence equation follows automatically. The first is where
\begin{equation}
F_{\mu\nu} = 2\partial_{[\mu} K_{\nu]}
\end{equation}
for a Killing vector $K_\mu$, and the second is where there is a self-duality condition
\begin{equation}
F_{\mu\nu} = {\phi_{\mu\nu}}^{\rho\sigma}F_{\rho\sigma}
\end{equation}
 for some covariantly constant four form $\phi$ on the manifold. (The Killing vector-defined field strength may also  satisfy a self-duality condition of course, or have this imposed.) We will study examples of both situations below. 

In the first case, where we construct $F$ from the Killing vector $K$, we note that $K^\mu F_{\mu\nu}$ is closed. This follows from application of Cartan's formula for the Lie derivative:
\begin{equation}
 \mathcal{L}_K F = i_K d F + d i_K F  \,
\end{equation}
which means for closed $F$ and $K$ Killing then $d i_K  F=0$.
Locally we can solve this condition to write  $K^\mu F_{\mu\nu}$ an exact form so that
\begin{equation}
\label{sigma}
 K^\mu F_{\mu\nu}= \partial_\nu \sigma\, .
\end{equation}
A solution to this (up to the addition of an arbitrary constant) is
\begin{equation}
\label{sigmaKsq}
\sigma= K^\mu K_\mu\, .
\end{equation}

In four dimensions, an analysis of the action of duality transformations \cite{Alawadhi:2019urr,Mars:2001gd}  shows that this is the same $\sigma$ as in equation \eqref{CeqCF}. See also  \cite{Walker:1970abc,Hughston:1972xyz}  where this was seen earlier using the spinor formalism in Type D spacetimes and more recently  \cite{Frolov:2017kze,Mason:2010zzc} for related work in higher dimensions.

The formula \eqref{CeqCF} naturally implies some conditions on the spacetime. Write the eigenvector equation for 
$F$ as
\begin{equation}
\label{evaF}
{F_\mu}^\nu k_\nu= \lambda_k k_\mu \, ,
\end{equation}
for eigenvector $k_\mu$ and eigenvalue $\lambda_k$. Note that the eigenvectors are necessarily null. Since $F$ is antisymmetric, its eigenvalues form into pairs with opposite signs (and one zero eigenvalue if $D$ is odd). Then  \eqref{CeqCF} implies that 
\begin{equation}
\label{Ceva}
C_{\mu\nu\rho\sigma} k^\nu  k^\sigma =\Lambda k_\mu k_\rho\, 
\end{equation}
with 
\begin{equation}
\Lambda = \frac{3}{ 2(D-1)(D-2)}\Big[ (D-1)(D-4)\lambda_k^2 + F_{\alpha\beta}F^{\beta\alpha}\Big]\, . 
\end{equation}
This implies that the eigenvector $k_\mu$ is a principal null direction of the Weyl tensor.  
Equation \eqref{CeqCF} also implies that 
\begin{equation}
\label{typeD_moreD}
k_{[\alpha}C_{\mu]\nu[\rho\sigma} k_{\pi]}k^\nu =0 \, .
\end{equation}
and similarly for a second eigenvector $l_\mu$ with a different eigenvalue. Thus generically there are two principal null directions satisfying \eqref{typeD_moreD} which implies that the spacetime satisfies the conditions for falling within the type D  class in the appropriate higher-dimensional classification  \cite{Ortaggio:2012jd} (see also the overview \cite{Reall:2011ys}). A special case occurs when the eigenvectors are not independent and so there is only one principal null direction. This occurs when $F$ obeys a self-duality relation in which case the two eigenvectors are identical. The spacetime is then said to fall into the type II class.


\subsection{Invariants}
Equation \eqref{CeqCF} implies that scalar invariants constructed from the Weyl tensor (and its dual where there is a suitable four-form on the manifold) are functions of the traces
\begin{equation}
(F^n) = {F^{\mu_1}}_{\mu_2} {F^{\mu_2}}_{\mu_3}\dots {F^{\mu_n}}_{\mu_1} \, .
\end{equation}
We will use a similar notation for traces of products of Weyl tensors (or their duals), for example
\begin{equation}
(C^3) ={C^{\mu\nu}}_{\rho\pi} {C^{\rho\pi} }_{\alpha\beta}{C^{\alpha\beta} }_{\mu\nu}  \, .
\end{equation}
In the four-dimensional case, then,
\begin{align}
\label{trExs}
(C^2)  &= {1\over \sigma^2}\Big(\,  {9\over 4} (F^2)^2 - 3 (F^4) \Big)\,  , \nonumber     \\ 
(CC^*)  &= {1\over \sigma^2}\Big(\,  (F^2)(FF^*) +2 (F^3F^*) \Big)\, ,    \nonumber \\ 
(C^3) &= {1\over \sigma^3}\Big(  -{61\over 2}(F^6) +{201\over 8}(F^2)(F^4) - {41\over 8}(F^2)^3  \Big) \,  \nonumber\\ 
(C^2C^*) &= {1\over \sigma^3}\Big(  {21\over 2}(F^3(F^3)^*) -(F^2)(F^3F^*) +\big(- {5\over 2}(F^2)^2+2(F^4)\big)(FF^*)  +{11\over 2}(F^5F^*)\Big) \, ,
\end{align}
and so on. We define $F^*_{\mu\nu}={1\over 2}\epsilon_{\mu\nu\rho\sigma}F^{\rho\sigma}$, and ${(F^3)^*}_{\mu\nu}={1\over 2} \epsilon_{\mu\nu\rho\sigma}F^{\rho\alpha}F_{\alpha\beta}F^{\beta\sigma}$ in the above.  For four-dimensional vacuum spacetimes, the invariants \eqref{trExs} form an  independent basis (under algebraic relationships) for the set of scalar invariants \cite{Lim:2004xx}.
In addition, traces of higher powers of $F$ are related to those of lower powers by the recursion relation
\begin{equation}
\label{4Drecrel}
(F^{2n}) ={1\over 2} (F^{2n-2})(F^2) -{1\over 8}\Big( (F^2)^2 - 2 (F^4)\Big) (F^{2n-4}) \, .
\end{equation}
Using this, one can express any $(F^n)$ in terms of  $(F^2)$ and $(F^4)$, and hence all scalar curvature invariants in the vacuum case are functions of these two traces.  There are analogous results in $D> 4$.

If one block-diagonalises $F$ as
\begin{equation}
\label{Fmatrix}
F = \begin{pmatrix}
0& x & 0 & 0 \\
-x & 0 & 0 &0 \\
0& 0 & 0 & y \\
0 & 0 &-y &0 
\end{pmatrix}
 \, .
\end{equation}
then $(F^2) = -2(x^2+y^2)$ and $(F^4) = 2(x^4+y^4)$ and the invariants are functions of these combinations. If $x$ and $y$ are proportional then $(F^2)$ is the only independent function. 

Similar arguments to those above apply to express the Weyl-NP scalars in terms of the independent Maxwell-NP scalars, depending on the dimension. We now give some examples to illustrate the above.


 \subsubsection{Taub-NUT}\label{Taub-NUT}

The Taub-NUT metric can be written in the form \cite{Ortin:2015hya}
\begin{equation}
\label{Taub-NUT metric}
  ds^2=-f(r)(dt-2n\cos{\theta}d\phi)^2+f(r)^{-1}dr^2+(r^2+n^2)d\Omega^2_2\, 
 \end{equation}
with
\begin{equation}
  f(r)=\frac{r^2-2mr-n^2}{r^2+n^2}\, .
\end{equation}
From the Killing vector $K^\mu=(1,0,0,0)$ define the Maxwell field $F_{\mu\nu}=2\partial_{[\mu}K_{\nu]}$. Then we have the Weyl doubling formula \cite{Alawadhi:2019urr}
\begin{equation}
\label{TNWeyl}
C^+_{\mu\nu\rho\pi} = {1\over\sigma^+}\, C_{\mu\nu\rho\pi} [F^+] \, ,
\end{equation}
with $\sigma^+ =-\frac{m+in}{r+in}=:-\frac{m_+}{r_+}$.
The eigenvalues of $F^+$ are $\frac{m_+}{r_+^2}(1,1,-1,-1)$ with corresponding eigenvectors
\begin{align}
\label{TNnulltetrad}
m^\mu &= \frac{1}{\sin{\theta}\sqrt{2(r^2+n^2}} \Big(-2n\cos{\theta},0,-i,0 \Big)\, , \quad 
l^\mu = \frac{1}{\sqrt{\lambda}}   \Big(\lambda,1,0,0 \Big)\, ,   \nonumber \\
\bar m^\mu &= \frac{1}{\sin{\theta}\sqrt{2(r^2+n^2}}  \Big(-2n\cos{\theta},0,i,0 \Big)\, ,  \quad
n^\mu =\frac{1}{\sqrt{\lambda}}   \Big(-\lambda,1,0,0 \Big)\, , 
\end{align}
with $\lambda=(r^2+n^2)/(n^2+2mr-r^2)$. These form a null tetrad, with $l_\mu n^\mu=-1,m_\mu\bar m^\mu=1$ and the remaining inner products zero.

As $F^+$ has repeated eigenvalues, the only independent trace is
\begin{equation}
\label{TNMaxn}
(F^+F^+)= -4\,\frac{m_+^2}{r_+^4}   \,   
\end{equation}
leading to
\begin{equation}
\label{TNWeyln}
(C^+C^+)= 24{m_+^2\over r_+^6}\, .
\end{equation}
All other scalar invariants are functions of this expression and its conjugate. 

The formula \eqref{TNWeyl} implies corresponding relationships amongst the NP scalars. In this case, for example, the Maxwell-NP scalars are $\phi_0= F^+_{\mu\nu}l^\mu m^\nu= 0, \phi_2 = F^+_{\mu\nu}\bar m^\mu n^\nu=0$ and 
\begin{equation}
\label{TNspinor}
 \phi_1= \frac{1}{2}F^+_{\mu\nu}( l^\mu n^\nu + \bar m^\mu m^\nu)= \frac{m_+}{r_+^2}    \, .
\end{equation}
Correspondingly, the only non-vanishing Weyl-NP scalar is 
\begin{equation}
\label{TNspinor2}
 \Psi_2= C^+_{\mu\nu\rho\pi}  l^\mu m^\nu \bar m^\rho n^\pi=    -\frac{m_+}{r_+^3} =\frac{1}{\sigma_+}\phi_1^2   \, .
\end{equation}
This simple form of the Weyl doubling formula can also be seen directly from the spinor formulation (c.f. \cite{Luna:2018dpt}).


\subsubsection{Plebanski-Demianski}
The  general vacuum type D solution with vanishing cosmological constant \cite{Plebanski:1976gy}, as given in  \cite{Luna:2018dpt}, is
\begin{equation}
\label{typeDmetric}
ds^2 = {1\over(1-pq)^2}  \Bigg[2 i (du+q^2dv) dp - 2(du-p^2dv)dq +  \frac{P(p)}{p^2+q^2}\,(du+q^2dv)^2
-  \frac{Q(q)}{p^2+q^2}\,(du-p^2dv)^2  \Bigg] \, ,
\end{equation}
with
\begin{align}
P(p) =  \gamma (1-p^4) +2 n p- \epsilon p^2 +2 m p^3 \,, \nonumber \\
Q(q) =  \gamma (1-q^4) - 2m q + \epsilon q^2 -2 n q^3 \,,
\end{align}
where the parameters $m,n,\gamma,\epsilon$ are related to the mass, NUT charge, angular momentum and acceleration (c.f. \cite{Griffiths:2005qp}).
The self-dual part of the Maxwell two form is given by
\begin{equation}
\label{PDMax}
F^+ =      {(m-in)\over2(p+iq)^2}  \Big(  i (du+q^2dv) dp + (-du+p^2dv)dq \Big), 
\end{equation}
with the anti-self dual part given by the complex conjugate.
One has the Weyl doubling formula
\begin{equation}
\label{PDWeyl}
C^+_{\mu\nu\rho\pi} = {1\over\sigma^+}\, C_{\mu\nu\rho\pi} [F^+] \, ,
\end{equation}
with $\sigma^+ =(m-in)(1-pq)^4/(4i(p+iq)$.
Then
\begin{equation}
\label{PDMaxn}
(F^2)= \Big({(m-in)(1-pq)^2\over (p+iq)^2}\Big)^2 , \quad  (F^4)=  {1\over 4} \Big({(m-in)(1-pq)^2\over (p+iq)^2}\Big)^4    \,   .
\end{equation}
These are not independent as the eigenvalues of $F^+$ are repeated - they are $\pm (m-in)(1-pq)^2/2(p+iq)^2$ twice, and hence 
\begin{equation}
\label{PDWeyln}
(C^+C^+)= -{24(m-in)^2(1-pq)^6\over(p+iq)^6}
\end{equation}
is the only independent curvature invariant involving $C^+$. Invariants involving $C$ and $C^*$ can be written in terms of this and its conjugate.

 
 \subsubsection{Eguchi-Hanson}
The Eguchi-Hanson metric is a vacuum solution with self-dual Weyl curvature. It is given by
\begin{equation}
\label{EHdoubleKS}
ds^2= 2dudv-2dXdY+\frac{\lambda}{(uv-XY)^3}(vdu-XdY)^2\, ,
\end{equation}
with coordinates $(u,v,X,Y)$ and constant $\lambda$. The single-copy (self-dual) Maxwell tensor is 
\begin{equation}
\label{EHMax}
F =  \frac{2 \lambda}{(uv-XY)^3} \Big( (u v+XY)(du\wedge dv - dX\wedge dY) - 2vY du\wedge dX + 2uX dv\wedge dY\Big) \, .
\end{equation}
This is the single copy tensor discussed in \cite{Luna:2018dpt}, rather than the ``mixed'' version of  \cite{Berman:2018hwd}.   With the Killing vector $K^\mu = (u,-v,-X,Y)$ this is given by $F_{\mu\nu}=(2\partial_{[\mu} K_{\nu]})^+$.
The Weyl curvature is then given by
\begin{equation}
\label{WeylEH}
C_{\mu\nu\rho\sigma} =  \frac{1}{\sigma^+} \, C_{\mu\nu\rho\sigma} [F ]   \,  ,
\end{equation}
with $\sigma^+ = -\frac{\lambda}{(uv-XY)}$ and $F$ the Maxwell field \eqref{EHMax}.

As in the case above, the eigenvalues of $F$ are repeated - here they are $\pm m/(uv-XY)^2$ twice. Thus the only independent trace of $F^n$ is
\begin{equation}
\label{EHMaxn}
(F^2)=  {4\lambda^2\over (uv-XY)^4} 
\end{equation}
and hence the only independent curvature invariant is
\begin{equation}
\label{PDWeyln2}
(C^2)= {24\lambda^2 \over(uv-XY)^6}\, .
\end{equation}
%

 
 \subsubsection{Singly rotating Myers-Perry}
 \label{MPsubsection}
 
As an example in $D>4$, consider the singly-rotating Myers-Perry metric in the form \cite{Elvang:2003mj}
\begin{align}
\label{MP}
  ds^2 =& -\frac{f(x)}{f(y)} \left(dt+
     R\sqrt{\nu} (1 + y) d\psi\right)^2       \nonumber \\
   & +\frac{R^2}{(x-y)^2}
   \left[ -f(x) \left( g(y) d\psi^2 +
   \frac{f(y)}{g(y)} dy^2 \right)
   + f(y)^2 \left( \frac{dx^2}{g(x)} 
   + \frac{g(x)}{f(x)}d\phi^2\right)\right]  \,, 
\end{align}
with
\begin{equation}
  f(\xi) = 1 - \xi \, ,
\qquad  g(\xi) = (1 - \xi^2)(1-\nu \xi) \, .
\end{equation}
From the Killing vector $K^\mu=\partial/\partial t$ form the Maxwell field strength $F_{\mu\nu} = 2\partial_{[\mu}K_{\nu]}$. Then one can check that
the Weyl tensor is given by the doubling formula
\begin{equation}
\label{MPWeyl}
C_{\mu\nu\rho\pi} = {1\over\sigma}\, C_{\mu\nu\rho\pi} [F] \, ,
\end{equation}
with $\sigma= (y-1)/(x-y)$. This satisfies \eqref{sigma}. One can ask how $K_\mu$ might be related to the standard single-copy gauge potential that arises from the Kerr-Schild form of the metric. In the notation of \cite{Frolov:2017kze}, section (E.3), the single-copy potential for the Myers-Perry metrics is $H l_\mu$, and it differs from the contravariant vector $K_\mu$ obtained from the Killing vector $\partial/\partial\tau$ by a gauge transformation with parameter $\tau$. Thus the two potentials give the same single-copy Maxwell tensor in  Kerr-Schild coordinates. When one transforms to the coordinates used in \eqref{MP} above, one obtains that metric, but with the function $g(\xi)$ having no term linear in $\xi$, i.e.,  with $g(\xi) = 1-\xi^2+\nu \xi^3$ \cite{Emparan:2001wn}. The coefficient of the linear term is purely kinematical \cite{Plebanski:1976gy} and does not affect the vanishing of the Ricci tensor. However, it does appear in the Weyl tensor, and its presence thus affects whether there is a Weyl doubling formula or not. For the function $g[\xi)=1-A \nu\xi -\xi^2+\nu \xi^3$ it can be checked that only for $A=1$ is there a Weyl doubling formula, which is \eqref{MPWeyl}.

From \eqref{Dhigher} with $D=5$, the $(C^2)$ and $(C^3)$  invariants are given by the formul\ae
\begin{align}
\label{trMP}
(C^2)  &= {1\over \sigma^2}\Big(\,  {15\over 8} (F^2)^2 - {3\over 2} (F^4) \Big)\,  ,  \nonumber    \\ 
(C^3) &= {1\over \sigma^3}\Big(  -{23\over 2}(F^6) +{81\over 8}(F^2)(F^4) - {41\over 16}(F^2)^3  \Big) \, .
\end{align}
The recursion relation \eqref{4Drecrel} also applies in five dimensions  so that the only independent traces are again $(F^2)$ and $(F^4)$. Here they are
\begin{align}
\label{trMPMax}
(F^2)  &= {2(-1+L^2(1+2x))(x-y)^3\over R^2(-1+y)^3}\,  ,      \\ \nonumber
(F^3) &={2(1-2L^2x+L^4(1+2x+2x^2))(x-y)^6\over R^4(-1+y)^6}\, .
\end{align}
The Maxwell field strength here has four distinct eigenvalues - two different pairs with opposite signs, and one zero.

Turning to the Weyl-NP scalars, the classification of  spacetimes in general dimensions via properties of the Weyl tensor has been discussed by Coley, Milson, Pravda and Pravdova (CMPP) in
\cite{Milson:2004jx,Coley:2004jv} (see also \cite{DeSmet:2002fv,Godazgar:2010ks}). More recently, the relationship between these classifications has been discussed and compared to a  spinor-based analysis  in \cite{Monteiro:2018xev}, who  show that in five dimensions the CMPP and spinor approaches are equivalent. The classification of Maxwell fields is also described there. 

We first define a null pentad of five  vectors $n^\mu, l^\mu, m_i^\mu$, ($i=1,2,3$), with
\begin{equation}
\label{nullpentad}
n^\mu l_\mu=-1, \;\; m_i^\mu m_{j\mu}= \delta_{ij}\, 
\end{equation}
and all other inner products zero. A convenient set is defined in the Appendix.
Now expand a five-dimensional Maxwell field as
\begin{equation}
\label{MaxExp5D}
F_{\mu\nu} =F_{01} n_{[\mu}l_{\nu]}   + \hat F_{0i} n_{[\mu}m_{\nu]i} + \tilde F_{1i} l_{[\mu}m_{\nu]i} + F_{ij} m_{i[\mu}m_{\nu]j} \, .
\end{equation}
There is an analogous expansion for the Weyl tensor. In this type D case the terms with non-zero weights vanish using our pentad choice and we have \cite{Milson:2004jx,Coley:2004jv}
\begin{equation}
\label{WeylExp5D}
C_{\mu\nu\rho\pi} =4C_{0101} n_{\{\mu}l_{\nu} n_{\rho}l_{\pi\}}  +  C_{01ij} n_{\{\mu}l_{\nu} m^i_{\rho}m^j_{\pi\}}  +  8C_{0i1j} n_{\{\mu}m^i_{\nu} l_{\rho}m^j_{\pi\}}  +  C_{ijkl} n_{\{\mu}m^i_{\nu} l_{\rho}m^j_{\pi\}} \, ,
\end{equation}
with the notation $T_{\{\mu\nu\rho\pi\}}:= {1\over 2} (T_{[\mu\nu][\rho\pi]} + T_{[\rho\pi][\mu\nu]} )$ for a tensor $T_{\mu\nu\rho\pi}$.

The Weyl doubling relationship \eqref{CeqCF} implies that the Weyl coefficients, such as those in \eqref{WeylExp5D}, are given in terms of the analogous Maxwell coefficients, which are given in five dimensions  by \eqref{MaxExp5D}. We also saw a simple  four-dimensional example of this for the Taub-NUT metric in subsection  \ref{Taub-NUT} above. In five dimensions we find, for example,
\begin{equation}
\label{NPmap}
C_{0101} ={1\over\sigma} \Big(  {3\over 16}F_{01}^2  - {1\over 8} \hat F_{0i}\tilde F_{1i} - {1\over 8} F_{ij}F^{ij}\Big) \, .
\end{equation}
For the Myers-Perry metric \eqref{MP} we find the non-zero Maxwell-NP scalars
\begin{equation}
\label{MPMaxScalars}
F_{01} = 2 {(x-y)\over R(-1+y)}\sqrt{{(-1+L^2 x)(x-y)\over (-1+y)}  }\, , \;\; 
F_{31}=-F_{13}= {L(x-y)\over R(1-y)}\sqrt{{(1+x)(x-y)\over(1-y)}}\, 
\end{equation}
and the non-zero independent Weyl-NP scalars
\begin{equation}
\label{MPWeylScalars}
C_{0101} = {(-3+L^2(1+4x))(x-y)^2\over 4R^2(1-y)^2}\, , \;\; 
C_{0131} = -C_{0113}= -{L^2(x-y)^4(1+x)(-1+L^2x) \over R^4(1-y)^4}\, .
\end{equation}
(The $C_{ijkl}$ are not independent quantities in five dimensions (c.f. also \cite{Monteiro:2018xev} eqn. (5.30)). One can then check that for the Myers-Perry metric \eqref{NPmap} is satisfied using \eqref{MPMaxScalars} and \eqref{MPWeylScalars} and the vanishing of $\hat F_{0i}$ and $\tilde F_{1i}$.


\section{The Gibbons-Hawking metrics}
\label{sec:GH}
 
We will now consider the extension of the Eguchi-Hanson case where there is a self-duality condition on the fields. Here we will find a generalisation of the Weyl doubling formula  \eqref{CeqCF}, where the additional terms are also given in terms of the derivative field strength $F$.  The Gibbons-Hawking metric is given by
\begin{equation}
\label{GHmetric}
ds^2 = {1\over V} \left( dx^4 + A_i dx^i\right)^2 + V dx^i dx_i \, ,
\end{equation}
where the fields $V, A^i$ are functions of the spatial coordinates  $x^i, i=1,2,3$, and are related by
\begin{equation}
\label{SDVA}
\nabla V = \nabla\times A\, .
\end{equation}
This equation implies $V$ is harmonic. It may thus be solved by a superposition of harmonic functions with arbitrary centres. The two centre solution can be shown to be equivalent to the Eguchi-Hanson solution, after a coordinate transformation.
From the Killing vector $K^\mu=(0,0,0,1)$ we can form the anti-self-dual  field strength
\begin{equation}
\label{SDVA2}
F_{\mu\nu}  = 2\partial_{[\mu}K_{\nu]} \, ,
\end{equation}
satisfying $F_{\mu\nu}^*=\frac{1}{2}\epsilon_{\mu\nu\rho\sigma}F^{\rho\sigma}=-F_{\mu\nu}$ with $\sigma=1/V$ via \eqref{sigma}. 
We find in this case that a simple Weyl doubling formula of the form of equation \eqref{CeqCF} does not hold.
To explore this further, note that as the Weyl curvature is anti-self-dual, it has only five independent components, which may be taken to be the components of the symmetric traceless matrix $3\times 3$ matrix $C_{i4j4}$. One can express this using the following three anti-self-dual two forms. First note that the vierbein one-forms are given by
\begin{align}
\label{vierbeinGH}
e^4 &={1\over \sqrt{V} }\big(dx^4 + A_i dx^i\big)   \, , \\ \nonumber
e^i  &= \sqrt{V} dx^i \, .
\end{align}
Then the two forms are given by
\begin{equation}
\label{SDtwoformsGH}
\Sigma^i = e^4 e^i + {1\over 2} \epsilon^{ijk}e^j e^k \, .
\end{equation}
We now solve the following equation for $\hat \Omega^i$:
\begin{equation}
\label{hatAGH}
d \Sigma^i +\epsilon^{ijk}\hat \Omega^j \Sigma^k  =0\, .
\end{equation}
This implies that $\hat \Omega^i $ is given by the anti-self-dual part of the spin connection, $\omega^\mu{}_\nu$ and thus
\begin{equation}
\hat{\Omega}^i= \omega^{i4}- \frac{1}{2} \epsilon^i {}_{jk} \omega^{jk} \, .
\end{equation} 
The  curvature can then be constructed directly from the curvature of $\hat \Omega^i$ and is given by 
\begin{equation}
\label{curvGH}
\hat C^i =d\hat \Omega^i+{1\over 2} \epsilon^{ijk}\hat \Omega^j \hat \Omega^k  =  \hat C^{ij}\Sigma^j \, ,
\end{equation}
where we have introduced $\hat C^{ij}$ encoding the Weyl tensor.
We  now use Cartan's first structure equation to calculate the connection one forms for the vierbeins as follows
\begin{equation}
\omega^{4i}= V^{-3/2}(- \frac{1}{2} \partial^i V e^4 + \partial^{[i} A^{j]} e_j) \, , \qquad \omega^{ij} =V^{-3/2} ( \frac{1}{2} \partial^j V  e^i - \partial^{[i} A^{j]} e^4)\, .
\end{equation}
Then, expanding both the Maxwell  field strength and the  curvature in terms of the two forms $\Sigma^i$ we find the following doubling formula
\begin{align}
\label{SDCGH}
F &=\alpha_i  \Sigma^i   \, , \\ \nonumber
\hat C^i  &= \big(V \alpha_i\alpha_j - \partial_i\alpha_j \big)^T  \, \Sigma^j \, ,
\end{align}
where the superscript $T$ means to take the traceless part of the expression within the brackets and
\begin{equation}
\label{alphaGH}
\alpha_i = {1\over V^2} \, \partial_i V  \, .
\end{equation}
In terms of the Weyl tensor we have the relation $\hat C_{ij} = -2 C_{i4j4}$, with the other components of the Weyl tensor  related to these by the anti-self-duality condition. Equation \eqref{SDCGH} may be viewed a a generalised form of Weyl doubling, where the quadratic, algebraic terms involving the gauge field are supplemented by terms depending on the derivatives of the gauge field. Thus we see that the general class of gravitational instantons in four dimensions satisfies a generalised Weyl doubling formula. 

An explicitly covariant version of this formula can be found as follows. There is the identity
\begin{equation}
\label{GHKVeqn}
C_{\mu\nu\rho\sigma} = -2V\Big( K_{\mu}K^\lambda C_{\nu\lambda\rho\sigma} + K_{\rho}K^\lambda C_{\sigma\lambda\mu\nu}
-2 g_{\rho\mu}C_{\nu\lambda\pi\sigma}K^\lambda K^\pi\Big)_{[\mu\nu][\rho\sigma]}  
\, 
\end{equation}
where the notation $[\mu\nu][\rho\sigma]$ means to antisymmetrise the expression within the preceding brackets in $\mu,\nu$ and separately in $\rho,\sigma$.  The above relation follows from the fact that $C^*=-C$ and $K^2=-\sigma=\frac{1}{V}$. To see this, define the three expressions on the right-hand side of \eqref{GHKVeqn}, including the antisymmetrisations, as $C_L+C_R+C_g$. Then, taking the (left) dual of each by contracting with  $\frac{1}{2}\epsilon^{\alpha\beta\mu\nu}$, one finds that $(C_L+C_R+C_g)^*=(-C-C_L)+(C_R)+(C_L-C_R)=-C=C^*$.

Equation  \eqref{GHKVeqn} is  equivalent to
\begin{equation}
\label{GHfinal}
C_{\mu\nu\rho\sigma} = V\Big( K_{[\mu}\nabla_{\nu]}F_{\rho\sigma} + K_{[\rho}\nabla_{\sigma]}F_{\mu\nu}\Big)^T\, ,
\end{equation}
where the notation $(...)^T$ means to subtract all traces. This can be re-expressed as the doubling formula
\begin{equation}
\label{GHfinal2}
C_{\mu\nu\rho\sigma} = V C_{\mu\nu\rho\sigma} [F]  +V\Big( 2\nabla_\rho(K_\mu F_{\nu\sigma}) -   \nabla_\mu(K_\nu F_{\rho\sigma})  -   \nabla_\rho(K_\sigma F_{\mu\nu}) - K_\mu\nabla_\nu F_{\rho\sigma} \Big)^T_{[\mu\nu][\rho\sigma]} 
\, .
\end{equation}
where the notation $(...)^T$ means to subtract all traces.

As the two-centred solution for the Gibbons-Hawking (GH) metric \eqref{GHmetric} is equivalent to the Eguchi-Hanson (EH) metric \eqref{EHdoubleKS} via a coordinate transformation one may wonder why there is a simple doubling formula for the EH metric described earlier, but not for the more general multi-centred GH metrics. This  special case can be understood by noting that for spherical polar coordinates the map from the EH metric to the two-centred GH metric (see \cite{Ghezelbash:2009we} for example) interchanges the periodic ``time'' coordinate and the 
azimuthal angle $\phi$. Mapping to the  two-centred GH case, the Killing vector which gives a simple Weyl doubling formula via the anti-self-dual part of the Maxwell field $2\partial_{[\mu}K_{\nu]}$ is then $K^\mu=\frac{\partial}{\partial\phi}$.
If one constructs the Maxwell field via the Killing vector $K^\mu=\frac{\partial}{\partial t}$ then one obtains the formul\ae\ in the analysis above for this particular potential. In the generic multi-centre case there is only the latter Killing vector, leading to the above analysis. (In the multi-centre case where the centres are all at different sites along the $z$-axis, the additional Killing vector is present of course - here we find evidence from a numerical analysis that the simpler Weyl doubling formula continues to hold.)


\section{An eight-dimensional example with Spin(7) holonomy}
\label{sec:SH}

The discussion above used features of the Gibbons-Hawking metrics which arise from the underlying self duality relation \eqref{SDVA}  which expresses the anti-self-duality of the field strength of the four vector gauge field $(V,A^i)$.  
In higher dimensions, manifolds of special holonomy are examples  where there are more novel duality conditions satisfied by the curvature, and two forms on the manifold can similarly be restricted to have duality properties determined by the canonical four-form defined by the special holonomy group (a pioneering paper on this is \cite{Corrigan:1982th}; see \cite{Gubser:2002mz} for a review relevant to string theory).
However, in general the formula \eqref{Dhigher} does not preserve duality properties. This means that if a two-form $F$ transforms in a certain representation of the holonomy group then this does not imply that $C[F]$ will as well (in each pair of indices). Furthermore,  projecting $C[F]$ onto appropriate representations does not in general preserve the algebraic symmetries needed to  relate it to a Weyl tensor. While these conditions may appear quite restrictive, we have found an example of an eight-dimensional manifold with Spin(7) holonomy  for which  Weyl doubling does work. We will use the conventions of \cite{Acharya:1997gp} in what follows. 
 
Begin with the metric \cite{Salur:2008pi}
\begin{equation}
\label{S7metric}
ds^2 = V^{-3/2} \big( dx^8 + A_i dx^i\big)^2 + V^{1/2} dx^i dx_i \, ,
\end{equation}
where here the indices $i,j,...$  run from $1$ to $7$,  and and $a,b,...$ and $\mu, \nu,...$ run from $1$ to $8$.  The fields $(A_i, V)$  are functions of the spatial coordinates  $x^i$ only. The spin(7) four-form $\phi_{abcd}$ has the following non-zero orthonormal frame components
\begin{align}
&[1256] =  [1278] =  [3456] =  [3478] =  [1357] =  
[2468] =  [1234] =  [5678] =  1    \, , \\ \nonumber
&[1368] =  [2457] =  [1458] =  [1467] =  [2358] =  [2367] = 
-1  \, ,
\end{align}
where $[abcd]$ means ${\phi}_{abcd}$. The four-form satisfies $\phi^2=4\phi+12$  and the projectors of a two form onto the
 \lq\lq self-dual" ${\bf 7}$ and  \lq\lq anti self-dual" ${\bf 21}$ representations of Spin(7) are given by
\begin{equation}
\label{projectors}
P_7 =  {1\over 4}(1+ {1\over 2}\phi), \qquad 
P_{21} =  {3\over 4}(1 - {1\over 6}\phi).
 \end{equation}
Define the acht-beins
\begin{align}
\label{achtbeinS7}
e^a &= (e^i,e^8)  \, , \nonumber \\ 
e^i  &=V^{1/4} \, dx^i    \, , \nonumber  \\ 
e^8 &=V^{-3/4} \, (dx^8 + A_i dx^i)\, .
\end{align}
Then the four form 
\begin{equation}
\label{fourform}
\phi =  {1\over 24} \phi_{abcd}e^ae^be^ce^d
 \end{equation}
 is closed provided that the fields $(A_i, V)$ satisfy the constraints  ($F_{ij} :=\partial_iA_j - \partial_j A_i$ and $V_i:=\partial_i V$)
\begin{equation}
\label{constraint}
V_{[i}\phi_{jklm]} - 2 F_{[ij}\phi_{klm]8} = 0\, .
 \end{equation}
These 21 equations can be solved in terms of 7 independent quantities. For example, choosing the independent variables to be $(F_{37}, F_{45}, F_{46}, F_{47}, F_{56}, F_{57}, F_{67})$ the other expressions are given by
\begin{align}
\label{Identities}
F_{12} &= F_{56}, F_{13} = F_{57}, F_{14} = -F_{67}, F_{15} = -F_{37}, F_{16} = F_{47}, F_{17} = -F_{46}, F_{23} = -F_{67}   \, , \nonumber \\ 
F_{24} &= -F_{57}, F_{25} = F_{47}, F_{26} = F_{37}, F_{27} = -F_{45}, F_{34} = F_{56}, F_{35} = -F_{46}, F_{36} = F_{45}   \, , \nonumber \\ 
V_1 &= -2F_{45}, V_2 = 2F_{46}, V_3 = 2F_{47}, V_4 = -2F_{37}, V_5 = 2F_{67}, V_6 = -2F_{57}, V_7 = 2F_{56}  \, .
\end{align}
It can be checked that $\phi$ is also covariantly constant when these conditions are met.  In fact, taking various linear combinations of the constraints
\eqref{constraint} one finds that $\partial_i \partial_j V=0$ for all $i,j$ so that $V$ is linear in the coordinates $x^i$ and the correction term is trivial. This also means that all the electric $V_i$ and magnetic $F_{ij}$ components of the gravitational field are constants.
Now consider the construction of the Maxwell field.  From the Killing vector $K^\mu=\delta^{\mu 8}$ we can again form the gauge  field strength
\begin{equation}
\label{KMax}
F_{\mu\nu}   = 2\partial_{[\mu}K_{\nu]} \, .
\end{equation}
This field is anti self-dual {\it i.e.,} in the {\bf 21} representation, as $P_7 F =0$. Define the anti self-dual two forms 
\begin{equation}
\label{ASDtwoformsS7}
\Sigma_-^{ij} =  P_{21} e^a \wedge e^b \, .
\end{equation}
These satisfy the orthonormality conditions 
\begin{equation}
\label{ASDortho}
\Sigma_-^{ab\mu\nu}\Sigma_{-cd\mu\nu} = 4{(P_{21})^{ab}}_{cd}, \qquad   \Sigma_{-}^{ab\mu\nu} \Sigma_{-ab\rho\sigma} =  {(P_{21})^{\mu\nu}}_{\rho\sigma}  \, .
\end{equation}
We can expand the gauge field two form in the basis of these as
\begin{equation}
\label{SCexpS7}
F_{\mu\nu}=  {1\over 2}\alpha_{ab}\, \Sigma^{ab}_{-\mu\nu}   \, ,
\end{equation}
with $\alpha_{ab}={1\over 2}\Sigma_{-ab}^{\mu\nu} F_{\mu\nu}$ which can be written as
\begin{equation}
\label{alphaS7}
  \alpha_{ij} = -{1\over 2V^2}\phi_{8ijk}V_k\, ,\qquad   \alpha_{i8} = -{3\over 2V^2} \, V_i  \, .
\end{equation}
The metric is Ricci-flat and it can be checked that the Weyl tensor $C_{\mu\nu\rho\sigma}$ satisfies $(P_7 C)_{\mu\nu\rho\sigma}=0$, with $P_7$ acting either on the first or second pair of indices of the Weyl tensor. The tangent space components of the Weyl tensor can be expressed purely in terms of the constants $V_i$. Let us then consider if these can be given as a doubling formula based on a  two form such as \eqref{SCexpS7} above. We will generalise this expression slightly and consider the ``twisted'' two form with components
\begin{equation}
\label{alphaS7ab}
  \alpha_{ij} =-{b\over 2V^2} \phi_{8ijk}V_k\, ,\qquad   \alpha_{i8} =  -{3a\over 2V^2}\, V_i  \, ,
\end{equation}
for some constants $a,b$. This two form will be anti self-dual if $a=b$. The natural expression to consider for a doubling formula is one based on 
\eqref{Dhigher} in eight dimensions, $C_{abcd}[\alpha]$ with $\alpha$ given by \eqref{alphaS7ab}. In four dimensions the formula \eqref{Dhigher} preserves duality - if $F_{\mu\nu}$ is self-dual or anti self-dual then $C_{\mu\nu\rho\sigma}[F]$ is also (in both pairs of indices). But this is not true in higher dimensions and  in this eight-dimensional case, in order to seek to match the Weyl tensor this expression must be projected onto the {\bf 21} on the left and right pairs of indices. Upon doing this we find that it is not possible to both preserve the duality conditions and the symmetry and trace properties of a Weyl-type tensor unless $a$ and $b$ satisfy the condition $23a^2-42 ab+27b^2=0$. We will write a solution of this as $a = \gamma b$ with $\gamma$ a particular complex number.
 In this case we find that there is a doubling  formula for the Weyl tensor
\begin{equation}
\label{weyl}
C_{abcd}=\lambda\, C^{21}_{abcd}[\alpha]   \, ,
\end{equation}
where the superscript $21$ on $C$ indicates that both pairs of indices are to be projected into the ${\bf 21}$ representation. The proportionality constant is $\lambda=-23 V^{3/2}/(32b^2(\gamma+1))$. The field $\alpha$ here is not anti self-dual as we have noted, but it can be checked that its anti self-dual part is proportional to the Maxwell tensor \eqref{SCexpS7}, and that the corresponding field strength $F_{\mu\nu}$ satisfies the Maxwell field equations.


\section{Reissner-Nordstrom and Born-Infeld}

In the sections above, we studied examples where the gauge field arises from intrinsic properties of the spacetime. We now turn to study situations where there is an {\it independent} gauge field defined on the spacetime, in order to see if there is a doubling relationship whereby 
the gauge fields in the gravity-gauge system give formul\ae\ for the full spacetime curvature - determining the Ricci tensor via the Einstein equations as usual but in addition fixing the Weyl tensor via  a doubling formula. We will call this ``self-doubling" for simplicity.

\subsection{Reissner-Nordstr\"om}
As a  first case, consider the Reissner-Nordstr\"om theory in $D$ dimensions, with metric
\begin{equation}
\label{RN}
ds^2 = -f(r) dt^2 + f^{-1}(r) dr^2 + r^2 d\Omega_{D-2}^2\, ,
\end{equation}
with 
\begin{equation}
\label{RN2}
f(r) = 1 - \frac{2M}{r^{D-3}} + \frac{Q^2}{r^{2(D-3)}}\, ,
\end{equation}
$d\Omega_{D-2}^2$ the metric on the $(D-2)$-dimensional sphere, and $M$ and $Q$ the mass and charge respectively. 

This theory is self-doubling in the sense that the gauge field already present in the theory provides the basis for the doubling. This can be seen as follows - the gauge field one form is
\begin{equation}
\label{RNsinglecopy}
A = -\frac{Q}{ r^{D-3}}\, dt \, ,
\end{equation}
satisfying the field equations ($F_{\mu\nu}=2\partial_{[\mu}A_{\nu]}$)
\begin{align}
\label{RNeqnsmotion}
\nabla^\mu F_{\mu\nu} & = 0\, , \nonumber  \\ 
R_{\mu\nu} -{1\over2}g_{\mu\nu}R &= \frac{D-2}{D-3} T_{\mu\nu} \, ,
\end{align}
with the usual stress tensor
\begin{equation}
\label{RNstresstensor}
T_{\mu\nu} = F_\mu\cdot F_\nu - \frac{1}{4}g_{\mu\nu}F\cdot F \, .
\end{equation}
We find the  Weyl doubling formula
\begin{equation}
\label{RNDC}
C_{\mu\nu\rho\sigma}  = \Lambda(r)\, C_{\mu\nu\rho\sigma}[F] \, ,
\end{equation}
where on the right-hand side of this  equation $C_{\mu\nu\rho\sigma}[F] $ is given by the formula \eqref{Dhigher} and the coefficient is
\begin{equation}
\label{RNcoeff}
\Lambda(r) = \frac{2(D-2)}{3(D-3)^2}\Bigg( (2D-5) - (D-1)\frac{M r^{D-3}}{Q^2}\Bigg)  \, .
\end{equation}
This result has been checked up to $D=10$ but there are no reasons why this would not hold for all dimensions given the structure of the metric.  Notice that one can take the extremal limit $Q\rightarrow M$ directly in the equations above. 


\subsection{Born-Infeld}
The discussion above may be generalised to the Born-Infeld theory in any dimension. The Lagrangian  is
\begin{equation}
\label{BILag}
L=   \sqrt{g} R + {1\over\lambda^2}\Big( \sqrt{g}-\sqrt{\vert \det{(g+\lambda F)} \vert}\Big) \, 
\end{equation}
with $F_{\mu\nu}$ the Maxwell field and $\lambda$ a constant. The solution with electric charge $Q$ in four dimension is given by (e.g.\cite{Rasheed:1997ns})
\begin{align}
\label{BI4D}
ds^2 & =   -\Big(1-{2 m(r)\over r^2} \Big)dt^2 + \Big(1-{2 m(r)\over r^2}\Big)^{-1}dr^2 + r^2 (dr^2+ d\Omega_2^2) \, , \nonumber  \\
F_{tr} &= -F_{rt}= {Q\over\sqrt{r^4+\lambda^2 Q^2} }\, ,
\end{align}
with the other components of $F$ vanishing. The function $m(r)$ is fixed by the metric equation of motion, and satisfies
\begin{equation}
\label{4Dm}
m'(r) =  \, {1\over \lambda^2}\Big(\sqrt{r^4+\lambda^2 Q^2} -r^2\Big)\, ,
\end{equation}
the solution of which is given by equation \eqref{4DmSol} below with $D=4$. We also define the tensor
\begin{equation}
\label{Gtensor}
G^{\mu\nu} =  -{2\over \sqrt{g}}{\partial L\over \partial F_{\mu\nu}}
\end{equation}
which satisfies the (non-linear in $F$) field equations $\nabla^\mu G_{\mu\nu}=0$. For the solution $F$ in \eqref{BI4D} this is given simply by
$G_{tr} = - G_{rt} ={ Q\over r^2}$ with other components vanishing. It is straightforward to confirm that the Weyl tensor satisfies the equation
\begin{equation}
\label{FWeylBI}
C_{\mu\nu\rho\sigma} = \Lambda_4(r) C_{\mu\nu\rho\sigma}[G]\, 
\end{equation}
with
\begin{equation}
\label{FWeylBILambda}
\Lambda_4(r) =-{2 r\over 3Q^2}\Big( 6 m(r) - 4 r m'(r) + r^2 m''(r)\Big)\, .
\end{equation}

This discussion is easily generalised to $D$ dimensions. The metric is as in \eqref{BI4D} with $d\Omega_2^2\rightarrow d\Omega_{D-2}^2$ with the function $m(r)$ satisfying 
\begin{equation}
\label{4Dm}
m'(r) =  \, {1\over \lambda^2}\Big(\sqrt{r^{2(D-2)}+\lambda^2 Q^2} -r^{D-2}\Big)\, .
\end{equation}
This is solved by
\begin{align}
\label{4DmSol}
\lambda^2\, m(r) =  \lambda^2\, m & + \frac{r^{D-1}}{D-1} - \frac{r \sqrt{\lambda^2Q^2+r^{2(D-2)}}}{D-1}\nonumber \\  & \quad + \frac{\lambda^2Q^2(D-2)}{(D-1)(D-3)r^{D-3}}\, \,{_2}F_1\Big[\frac{1}{2},\frac{D-3}{2(D-2)},1+\frac {D-3}{2(D-2)}, -\frac{\lambda^2Q^2}{r^{2(D-2)}}\Big] \, .
\end{align}
for constant $m$. Note that $m(r)\rightarrow m$ in the limit  $r\rightarrow0$.

One has the result 
\begin{equation}
\label{FWeylBID}
C_{\mu\nu\rho\sigma} = \Lambda(r) C_{\mu\nu\rho\sigma}[G]\, 
\end{equation}
with 
\begin{equation}
\label{FWeylBILambdaD}
\Lambda(r) =-{2 r^{D-3}\over 3 Q^2}\Big( (D-1)(D-2) m(r) - 2(D-2) r m'(r) + r^2 m''(r)\Big)\, 
\end{equation}
and the field $G$ in $D$ dimensions given by
\begin{equation}
\label{GD}
G_{tr} = - G_{rt} ={Q\over r^{D-2}}\, ,
\end{equation}
with other components vanishing. Notice that $G_{\mu\nu}$ is a function of the Maxwell field $F_{\mu\nu}$ via the relation \eqref{Gtensor}.
The result \eqref{FWeylBID} leads to curvature singularities as $r\rightarrow 0$ which are not present in the field $F_{\mu\nu}$.

This Born-Infeld solution reduces to the Reissner-Nordstr\"om model in the section above in the limit as $\lambda\rightarrow 0$, using the expansion
\begin{equation}
\label{BIexp}
\det{({{\delta^\mu}_\nu}+\lambda {{F^\mu}_\nu})}=  1-{1\over 2}\lambda^2(F^2) -{1\over 4}\lambda^4 \Big( (F^4) -{1\over2}(F^2)^2\Big) + o(\lambda^6)     \, ,
\end{equation}
where the brackets in $(F^2), (F^4)$ indicate matrix traces of powers of ${F^\mu}_\nu$. It can be checked that the coefficient in \eqref{FWeylBILambdaD} reduces in this limit to that in  \eqref{RNcoeff}.

The formula \eqref{FWeylBID} enables one to easily find invariants, for example
\begin{equation}
\label{CsqBI}
C^{\mu\nu\rho\sigma} C_{\mu\nu\rho\sigma} = \frac{4\,(D-3)}{(D-1)\,r^{2(D-1)}} \Big( (D-1)(D-2) m(r) - 2(D-2) r m'(r) + r^2 m''(r)\Big)^2 \, .
\end{equation}
Note that this diverges as $1/r^{2(D-1)}$ as $r\rightarrow 0$, although the non-zero Maxwell field strength component $F_{tr}= Q/\sqrt{r^{2D-4} +\lambda^2Q^2}$ does not. The divergence comes from the traces of the field $G_ {\mu\nu}$ which appear when squaring \eqref{FWeylBID}. (A discussion of the four dimensional Born-Infeld theory appeared recently \cite{Pasarin:2020qoa} and the $D=4$ version of \eqref{CsqBI} appeared there, although their equivalent of $m(r)$ satisfies a different equation.)


\section{Brane solutions}
\label{branes}

It is natural  to conjecture that BPS brane solutions in supergravity (we will follow the conventions of \cite{Stelle:1999ljt} in this section) might satisfy Weyl doubling.  The $(p+1)$-forms that minimally couple to the $p$-brane provide a potential from which one can construct a $(p+2)$-form field strength. This field strength can then be used in the formula for the Weyl curvature, as well as determining the Ricci curvature via the field equations and stress tensor. We  consider here the cases of $p$-branes where the scalar fields play no role.


\subsection{String in five dimensions}
A simple example is the string in five dimensions. The metric and two-form are given by
\begin{align}
\label{string5d}
ds^2 &= H^{-1}(r) \big(-dx_1^2 + dx_2^2 \big) + H^2(r) (dr^2 + r^2 \big(d\theta^2+{\rm sin}^2 \theta d\phi^2) \big) \, , \nonumber \\ 
B_{\mu\nu} &= \sqrt{3}\,\epsilon_{\mu\nu}H^{-1}(r) \, ,
\end{align}
with $H(r) = 1+k/r$ and indices $\mu,\nu=1,2$. The three-form field strength $F=dB$, obeys $\nabla^M F_{MNP}=0$ and the metric field equation
\begin{equation}
\label{stresstensor}
G_{MN} =\frac{1}{4}\Big( F_M\cdot F_N - \frac{1}{6}g_{MN}F\cdot F\Big) \, .
\end{equation}
For the $p$-forms $F$ discussed in the subsections below the equivalent equation is 
\begin{equation}
\label{stresstensorP}
G_{MN} =\frac{1}{2(p-1)!}\Big( F_M\cdot F_N - \frac{1}{2p}g_{MN}F\cdot F\Big) \, .
\end{equation}

We will now define the two-forms ${(F_\mu)}_{NP}={(K_\mu)}^M F_{MNP}$ ($\mu=1,2$) where the Killing vectors  $K_\mu$ correspond to translations in the $x^\mu$ directions in the string world sheet. Note that $K_1^2=H^{-1}(r)=-K_2^2$. We then find that the following Weyl doubling formula may be constructed from the three-form field strength $F$ and the two-form field strengths $F_\mu$
\begin{equation}
C_{MNPQ} = -\frac{(k+4r)}{6k} C_{MNPQ}[F] - \frac{r H(r)}{3k}\Big( -C_{MNPQ}[F_1] + C_{MNPQ}[F_2] \Big)\, .
\end{equation}
This can also be written  as
\begin{equation}
\label{5dstringDC}
{C_{MN}}^{PQ} = -\frac{1}{6}{C_{MN}}^{PQ}[F]  + \Sigma_1(r)  {T_{MN}}^{PQ}   \, ,
\end{equation}
where $ \Sigma_1(r)  = \frac{k }{r^3H^4}$. 
We note that $\Sigma_1(r)$ is proportional to the inverse of the volume of the transverse sphere. 
In this coordinate system all of the non-zero components of the Weyl tensor  take the form (up to sign)
${C_{MN}}^{MN}$.  The only non-zero components of ${C_{MN}}^{PQ}[F] $ and the tensor ${T_{MN}}^{PQ}$  are similarly  when $(P,Q)=(M,N)$ (or $(N,M)$). In this case the ${T_{MN}}^{PQ}$  are given by $(0,1,-1/2,-1,2)$ for $(MN)=(12,\mu 3,\mu \bar m, 3\bar m,45)$ respectively ($\bar m = 4,5$), and components related to these by the antisymmetry in $M,N$. 
Notice thus that the tensor $T$ vanishes when all its components are along the world-volume, and so from \eqref{5dstringDC} we see that on the world-volume there is a simple Weyl doubling formula, and this is corrected off the world volume by a tensor that takes a simple form.

In \cite{Lee:2018gxc} the Kerr-Schild formulation was investigated in the case where there is both a metric and a Kalb-Ramond field,  using doubled geometry. It was found that this involved two null vectors with the metric and $B$ field involving the symmetric and anti-symmetric product of these vectors. It is natural then to expect that a single copy in this case should involve two Maxwell gauge fields $A$ and $\bar A$. For the five-dimensional case under consideration here, one can see that the $B$ field
in \eqref{string5d} is given by
\begin{equation}
\label{BfieldDC}
B_{MN} = 2 \sqrt{3}\,H A_{[M}\bar A_{N]}   \, ,
\end{equation}
with 
\begin{align}
\label{BfieldSC2}
A_M &= (H^{-1},0,0,0,0)\, ,    \nonumber\\
\bar A_M &= (0,H^{-1},0,0,0) \,  .
\end{align}
This generates a formula for $F_{MNP}$ in terms of $A_M$ and $\bar A_M$. 

One avenue suggested by this work is to develop a Weyl doubling formula for DFT. Along these lines, one may view an expression of the formula \eqref{5dstringDC} as a Weyl ``doubling'' in terms of the two Maxwell fields in the formalism of \cite{Lee:2018gxc}. Inserting \eqref{BfieldDC} into \eqref{5dstringDC} would then give an expression for the Weyl tensor in terms of an expression quartic in fields and two derivatives. (Note that the expression \eqref{BfieldDC} is in the usual double copy where the vector fields in that formalism live in flat space.)


\subsection{M2 brane}

For the M2 brane in eleven dimensions, the metric and non-zero three-form potential components are given by
\begin{align}
\label{M2}
ds^2 &= H^{-2/3}(r) (-dx_1^2 + dx_2^2+ dx_3^2) + H^{1/3}(r) (dr^2 + r^2 d\Omega_{7}^2), \\ \nonumber
A_{\mu\nu\rho} &= \epsilon_{\mu\nu\rho}H^{-1}(r) \, ,
\end{align}
with $H(r) = 1+k/r^6$ and indices $\mu,\nu,...=1,2,3$, and $m,n,...=4,...,,11$ (with  $M,N,...=1,...,11)$. $r$ is the radial coordinate in
the eight-dimensional transverse space.
The four-form field strength $F=dA$ has non-zero components $F_{1234} = -6k H^{-2}/r^7$ and obeys $\nabla^M F_{MNPQ}=0$ and \eqref{stresstensorP} with $p=4$.

Again we define the tensors  ${(F_\mu)}_{MNP}={(K_\mu)}^Q F_{QMNP}$ where the Killing vectors  $K_\mu$ correspond to translations in the $x^\mu$ directions in the world-volume. The Weyl doubling formula can then be constructed as follows:
\begin{equation}
C_{MNPQ} = \frac{(k+4r^6)}{36k} C_{MNPQ}[F] - \frac{r^6 H(r)^{2/3}}{3k}\Big( -C_{MNPQ}[F_1] + C_{MNPQ}[F_2] + C_{MNPQ}[F_3] \Big) \, .
\end{equation}
This simplifies along the world-volume as in the string case above, with the analogue of \eqref{5dstringDC} being
\begin{equation}
\label{MDCfull}
{C_{MN}}^{PQ} = \frac{1}{36}{C_{MN}}^{PQ}[F]  + \Sigma_2(r)  {T_{MN}}^{PQ}   \, ,
\end{equation}
where here $ \Sigma_2(r)  = \frac{2k H^{-7/3}}{r^8}$ and the object ${T_{MN}}^{PQ} $ has non-zero components $(0,7,-1,-3,1)$ for $(PQ)=(MN) = (\mu\nu,\mu 4,\mu \bar m,4\bar m,\bar m\bar n)$ respectively, with $\bar m,\bar n=5,...,11$. Again we note that $\Sigma_2(r)$ is proportional to the inverse of the volume of the transverse sphere. 

Whilst a Kerr-Schild type formulation based on exceptional geometry (see \cite{Berman:2020tqn} for  a review) is not yet available, we note that for $SL(5)$ the $10$ representation reduces to $4+6$ in four dimensions, representing a vector and two-form which are expected to be fundamental to this. Thus one might expect that a single copy in this case might involve a Maxwell gauge field $A_M$ and two-form $A_{MN}$, with the $A_{MNP}$ field.


\subsection{D3 brane}

For the D3 brane in ten dimensions, the metric and non-zero four-form potential components are given by
\begin{align}
\label{D3}
ds^2 &= H^{-1/2}(r) (-dx_1^2 + dx_2^2+ dx_3^2+ dx_4^2) + H^{1/2}(r) (dr^2 + r^2 d\Omega_{5}^2), \\ \nonumber
A_{\mu\nu\rho\sigma} &= \epsilon_{\mu\nu\rho\sigma}H^{-1}(r) \, ,
\end{align}
with $H(r) = 1+k/r^4$ and indices $\mu,\nu...=1,2,3,4$, and $m,n,...=5...,,10$ (and  $M,N,...=1,...,10)$. $r$ is the radial coordinate in
the six-dimensional transverse space.
We will need the self-dual five-form field strength $F=\frac{1}{2}(dA+(dA)^*)$.  As above, we define  the tensors  ${(F_\mu)}_{MNPQ}={(K_\mu)}^R F_{RMNPQ}$ where the Killing vectors  $K_\mu$ correspond to translations in the $x^\mu$ directions in the world-volume. The Weyl doubling formula  is then found to be
\begin{equation}
C_{MNPQ} = \frac{r^4}{6k} C_{MNPQ}[F] - \frac{2r^4 H(r)^{1/2}}{3k}\Big( -C_{MNPQ}[F_1] + C_{MNPQ}[F_2]+ C_{MNPQ}[F_3] + C_{MNPQ}[F_4]  \Big)
\end{equation}
This also shows that the Weyl tensor vanishes for components along the world-volume. To see this, one has the equivalent expression 
\begin{equation}
\label{D3old}
{C_{MN}}^{PQ} = \Sigma_3(r)  {T_{MN}}^{PQ}   \, ,
\end{equation}
where  $ \Sigma_3(r)  = \frac{k H^{-5/2}}{r^6}$ and the object ${T_{MN}}^{PQ} $ has non-zero components $(0,5,-1,-4,2)$ for $(PQ)=(MN) = (\mu\nu,\mu 5,\mu \bar m,5\bar m,\bar m\bar n)$ respectively, with $\bar m,\bar n=6,...,10$. As in the cases above, $\Sigma_3(r)$ is proportional to the inverse of the volume of the transverse sphere. 
\\


\subsection{M5 brane}

There is a similar story for the M5 brane in eleven dimensions. The metric and non-zero 4-form field strength components are given by
\begin{align}
\label{M5}
ds^2 &= H^{-1/3}(r) \eta_{\mu\nu}dx^\mu dx^\nu + H^{2/3}(r) ( dr^2 + r^2 d\Omega_{4}^2 ),   \nonumber \\ 
F^{(4)}_{8\,9\,10\,11} &= 3k  \sin^3(\theta) \sin^2(\phi) \sin(\psi_1)\, ,
\end{align}
with $F^{(4)}$ antisymmetrised, $H(r) = 1+k/r^3$, world-volume coordinates $x^\mu$ ($\mu=1,...6)$, and transverse coordinates $r$ and spherical polars $(\theta,\phi,\psi_1,\psi_2)$. 

The fivebrane magnetically couples to the three form $C_3$ which means we will need to use the magnetic dual field strength given by the seven-form, $F={^*F^{(4)}}$.
Then define ${(F_\mu)}_{NPQRST}={(K_\mu)}^M F_{MNPQRST}$ ($\mu=1,2,...,6$) with the Killing vectors  $K_\mu$ corresponding to translations in the $x^\mu$ directions in the world-volume.  The Weyl doubling formula is then
\begin{align}
36 \, C_{MNPQ} = &-\frac{1}{60} C_{MNPQ}[F] +  \frac{r^3}{3k} C_{MNPQ}[F] \nonumber \\
  &-  \frac{r^3}{2k}H(r)^{1/3}\Big( -C_{MNPQ}[F_1] + C_{MNPQ}[F_2] + \dots + C_{MNPQ}[F_6] \Big) \, .
\end{align}

This is equivalent to
\begin{equation}
\label{M5DCfull}
36\,{C_{MN}}^{PQ} =- \frac{1}{60}{C_{MN}}^{PQ}[F]  + \Sigma_5(r) {T_{MN}}^{PQ}  \, ,
\end{equation}
where here $ \Sigma_5(r)  = \frac{18 k H^{-8/3}}{r^5}$ and the object ${T_{MN}}^{PQ} $ has non-zero components $(0,4,-1,-6,4)$ for $(PQ)=(MN) = (\mu\nu,\mu 7, \mu\, \bar m,7\bar m,\bar m\bar n)$ respectively, with $\bar m,\bar n=8,9,10,11$. We see again there is a simple doubling formula for the Weyl components along the world-volume directions and that $\Sigma_5(r)$ is proportional to the inverse of the volume of the transverse sphere. 

In all of  the cases above, for a brane with $V$-dimensional world-volume in $D$ dimensions, and $T=D-V$ transverse dimensions, the components of the tensor  ${T_{MN}}^{MN}$ 
are proportional to $(0,T-1,-1,-V,2V/(T-2))$ for $(PQ)=(MN) = (\mu\nu,\mu r, \mu\, \bar m,r\bar m,\bar m\bar n)$ respectively, in the notation used above. Given that the components along the world-volume vanish, this follows from the tracelessness condition on the Weyl tensor. The powers or $r$ and $H$ in the coefficients $\Sigma$ are  equal to $-T$ and that appearing in the inverse of the volume of the transverse sphere respectively. Similarly, the  powers of $r$ and $H$
in the doubling formul\ae\ above have common expressions: $r^{T-2}$ and the inverse of the power of $H$ which appears in the world-volume metric.


\section{Discussion}
\label{sec:disc}
We described Weyl doubling as the writing of the Weyl tensor (up to a scalar factor) in terms of a quadratic expression in an Abelian field strength obeying Maxwell's equations in a curved background. The curved background distinguishes these results from the usual double copy originating in scattering amplitudes and more recently classical solutions using the Kerr-Schild form. 
This phenomenon of Weyl doubling was found in a variety of solutions in different dimensions. The purely intrinsic case is where a Killing vector on the manifold is used to define a potential from which the field strength is derived. The Weyl tensor is given by the formula \eqref{Dhigher}. We showed that the  metrics for which the Weyl doubling formula \eqref{Dhigher}  applies fall into the Type D class in the general dimensional classification, so that this is a necessary condition.
It is also a sufficient condition in four dimensions \cite{Walker:1970abc,Hughston:1972xyz} (see also \cite{Luna:2018dpt}). But it does not appear to be a sufficient condition in higher dimensions - the  five-dimensional  Myers-Perry metrics \cite{Myers:2011yc}  are Type D \cite{DeSmet:2003kt} and while we showed that the singly-rotating solution has a Weyl doubling formula,  the general solution with two rotation parameters  does not appear to satisfy such a formula. We also investigated the BPS solution in six-dimensions studied in \cite{Chen:2006xh}, which is Type D but  does not appear to satisfy a doubling formula.

A second way to generate a two-form gauge field strength is if the spacetime admits a closed, covariantly constant four-form which may be used to define a self-dual (or anti-self-dual)  two form. These manifolds have special holonomy and the Maxwell field equations then follow from the duality and closure conditions.  In the four-dimensional case these are manifolds with $SU(2)$ holonomy. The Gibbons-Hawking metrics provide  a broad class of such metrics. We found that in this case the Weyl doubling formula has a correction term which is linear in the gauge field strength - equation \eqref{SDCGH} or equivalently \eqref{GHfinal2}. This generalisation of doubling could be studied further to see if it applies in other cases. It may also suggest generalisations of the double copy construction. 
A reader might wonder that  such a doubling formula is inevitable given the symmetries of the Weyl tensor. The Gibbons-Hawking case where there is a derivative correction provides a good counter example that demonstrates the non-triviality of the algebraic relation in the algebraic Weyl doubling formula.

In studying a generalisation to higher dimensions an issue arises in that a Weyl tensor constructed from \eqref{Dhigher} using a self-dual two-form is not in general  self-dual, unlike in four dimensions.  One might have expected that this might be resolved by using self-dual projection operators, but this will not preserve the algebraic symmetries of the tensor in general. Nevertheless we found an example in eight dimensions where a sort of twisted  doubling construction exists which expresses the Weyl tensor in terms of a spin(7) self-dual  two form. It would be interesting to explore if other examples exist.

A different mechanism for Weyl doubling is if the gauge field strength is an {\it additional} field defined on the spacetime, rather than being expressed using the metric and/or Killing vectors. The most natural example of this is the  Reissner-Nordstr\"om metric in $D$ dimensions. This works in the general case as well as the BPS limit, although perhaps one might have expected that such a construction, requiring the Weyl and Maxwell curvatures to be related, would require the BPS constraint linking the charge and mass. We showed that this discussion generalises to the charged Born-Infeld solution in $D$ dimensions.  A recent paper \cite{Pasarin:2020qoa} discussed how study of the Born-Infeld electrically charged solution might illuminate the investigation of stringy corrections of the double copy. It would be interesting to see if the doubling approach may provide insights into this. 

We then turned to study cases where the gauge field strength is a higher degree $p$-form, and analysed the associated brane solutions. Here it was found that a simple quadratic Weyl doubling formula holds using the $p$-form field strength and contractions of it with the world volume Killing vectors. Evaluating this gives a particularly simple Weyl doubling formula for the components of the Weyl tensor projected on to the world volume.
There is a variety of directions for further research, such as how does the inclusion of scalars, Kaluza-Klein reductions and supersymmetry impact the doubling construction. We have studied the Weyl tensor here but there will also be spinor analogues of our formul\ae\ in each dimension (c.f \cite{Monteiro:2018xev}). 

The central question that this paper implicitly poses is, what is the relation to the usual double copy? Can the formul\ae\ for brane solutions be applied to the usual single and double copy? Does the derivative corrected Gibbon-Hawking expression have a conventional double copy interpretation in terms of Maxwell fields in flat space? It was shown in  \cite{Kim:2019jwm} that doubled geometry clarifies the double copy construction for the point charge, relating this to the JNW solution \cite{Janis:1968zz}. As we noted in section \ref{branes}, the Kerr-Schild form in double field theory involves more than one gauge field - for example, two Maxwell fields in the case where there is a Kalb-Ramond field as well as the metric. For the string in five dimensions the $B$ field can indeed be simply constructed from two Maxwell gauge fields. This might also provide insights into exceptional geometry where such a Kerr-Schild formulation is not yet available.
One approach following these ideas is to use a DFT or EFT generalised Killing vector as the basis for generating gauge field strengths in the extended space to express the DFT equations. 

The phenomenon of Weyl doubling that we have explored here in numerous examples, relating gravity and Abelian gauge theory, reveals a  structure  that it would be interesting to develop further, and in particular to investigate its relationship to the double copy.

\section*{Acknowledgments}
DSB and BS are supported by the UK Science and Technology Facilities Council (STFC) with consolidated grant ST/L000415/1, String Theory, Gauge Theory and Duality. RA is supported by a student scholarship from the Ministry of Higher Education of the UAE. We thank Kanghoon Lee, Ricardo Montiero, Malcolm Perry and Chris White for discussions on various aspects of the double copy and classical solutions.

\section*{Appendix}

The pentad used for the analysis of the Myers-Perry metric in subsection \ref{MPsubsection} is based on the two null vectors $l^\mu=L^\mu_\pm$ satisfying the equation $l^\nu l^\rho C_{\mu\nu\rho[\pi}l_{\omega]}=0$ \cite{Pravda:2005kp}
\begin{equation}
\label{pentadnl}
L_\pm  = {1\over(x^2-1)(-1+L^2y)}\Bigg( {L^2x y - y + L^2 x +1 -2 L^2y\over x-y} R\partial_t -L\partial_\psi\Bigg) \pm \sqrt{{L^2x-1\over(x-y)(y-1)}}\Bigg(\partial_x+{y^2-1\over x^2-1}\partial_y\Bigg)
 \, .
\end{equation}
We take $l^\mu = L_+^\mu, n^\mu=L_-^\mu$ and choose the three unit norm vectors $m^\mu_i$ ($i=1,2,3$) to be
\begin{align}
\label{pentadmi}
m_1  &= \sqrt{ {(1+x)(x-y)(1+y)(-1+L^2y)^2\over (1+L^2)R^2(x-1)}} \Bigg(0,-{(x-1)(-1+L^2x)\over (y-1)(-1+L^2y)},-1,0,0\Bigg)\, ,  \nonumber  \\
m_2&= \sqrt{- {(x-y)^2\over (1+x)(-1+L^2x)(y-1)^2R^2}} \Bigg(0,0,0,-1,0\Bigg) \, ,  \nonumber\\ 
m_3&=  \sqrt{- {(x-y)^2\over (-1+x)(1+L^2)(y^2-1)R^2}}    \Bigg( {LR(x-1)(y+1)\over x-y},0,0,0,-1\Bigg)\, .
\end{align}
These vectors satisfy the conditions in and beneath \eqref{nullpentad}. This choice is convenient in this case as the non-zero weight Weyl-NP (and Maxwell-NP) components vanish directly using them.

\pagebreak

\bibliography{refs}

\end{document}